\documentclass[%
 reprint,
 amsmath,amssymb,
 aps,
]{revtex4-2}
\usepackage{cleveref}

\usepackage{graphicx}
\usepackage{dcolumn}
\usepackage{bm}
\usepackage{xcolor}

\begin{document}

\preprint{APS/123-QED}

\title{Emergent Inequalities in a Primitive Agent-Based Good-Exchange Model}

\author{Nirbhay Patil}
\affiliation{Laboratoire de Physique de l'\'Ecole Normale Supérieure, Paris 75005, France}
\affiliation{Chair of Econophysics and Complex Systems, École polytechnique, 91128 Palaiseau Cedex, France}
\author{Jean-Philippe Bouchaud}%
\affiliation{Capital Fund Management, Paris 75007, France
}%

\affiliation{Chair of Econophysics and Complex Systems, École polytechnique, 91128 Palaiseau Cedex, France}
\affiliation{Academie des Sciences, Paris 75006, France
}%
\date{\today}

\begin{abstract}
Rising inequalities around the globe bring into question our economic systems and the origin of such inequalities. Here we propose a toy agent-based model where each entity is simultaneously producing and consuming indivisible goods. We find that the system exhibits a non-trivial phase transition beyond which a market clearing equilibrium exists but becomes dynamically unreachable. When production capacity exceeds a threshold and adapts too slowly, some agents cannot sell all their goods. This leads to global price deflation and induces strong wealth inequalities, with the spontaneous separation of the population into a rich class and a poor class. We explore ways to alleviate poverty in this model and whether they have real life significance. 
\end{abstract}

\maketitle


\section{\label{sec:level1}Introduction}

Wealth and income inequalities have been at the forefront of political debate since time immemorial, and the direct causes of social unrest and economic upheavals, like the French and Russian revolutions, the cultural revolution in China, and many other major historical events. These inequalities have been recently brought into light for economists and other academics as well as policy makers by the influential work of Thomas Piketty \cite{piketty2014capital}. We currently live in a world of unprecedented inequalities - as an illustration: according to Saez and Zucman \cite{saez2016wealth}, the share of the America’s richest .01 percent nearly quadrupled since 1953, rising from 2.5 percent to close to 10 percent. Historical studies have shown that inequalities have overall grown with economic development -- an apparently paradoxical situation. 

What are the mechanisms leading to wealth inequalities, and in particular to the appearance of an ``oligarchy'', a phenomenon sometimes called wealth ``condensation''? Statistical models of income and wealth dynamics have a long history, starting with Champernowne \cite{champernowne1953model} and Angle \cite{angle1986surplus}, with a particular upsurge in the ``Econophysics'' literature since 2000 -- for recent reviews see e.g. \cite{chakrabarti2006econophysics, venkatasubramanian2017much,boghosian2019inequality,ribeiro2020income,greenberg2023twenty}, and, for economics papers \cite{benhabib2011distribution,gabaix2016dynamics}. It has been found that stochastic models of wealth exchange can indeed lead to skewed, Pareto-tailed distribution of wealth and even, in some cases, to extreme inequalities with the existence of an ``oligarchy'', defined as a finite {\it number} of individuals detaining a finite {\it fraction} of the total wealth, even when the total population becomes infinitely large. 

Whereas too much inequality is morally unfair and economically inefficient, perfect equality is difficult to achieve and its effects on global welfare unknown, making it difficult to decide what level of inequalities policy makers should target. If one thinks, for example, in terms of the classic Gini coefficient \footnote{The Gini coefficient $G \in [0,1]$ measures the average absolute income (or wealth) difference in a population, rescaled by (twice) the average wealth.} $G$, is there an optimal value of $G$ that we can rationally agree on? Should $G$ be less than $0.3$ like in Scandinavian countries, or is $0.35$ or even $0.47$ like in the US in 2022 still acceptable? Should this target be independent of the level of development of an economy? How does the level of inequality affect economic growth and global social welfare \cite{ostry2014redistribution} and vice versa?

In a recent note \cite{bouchaud2020much}, one of us argued that a possible mechanism for inequalities is productivity growth when the utility associated to the consumption of goods saturates. For full employment to be maintained, the consumption of low productivity, expensive goods must be bolstered. In a nutshell, the argument is based on the idea that if the productivity of a non-essential good is too low, its price is too high for any agent to afford in a strictly egalitarian individualist society. Hence, this good is not produced at all. But if at the same time the essential good is easy to produce while its utility saturates, only a fraction of the labour force is needed and output remains low. In this case, global welfare is higher when the Gini coefficient is non-zero, i.e. in the presence of some degree of wage inequalities. We will see below that some of the conclusions of Ref.~\cite{bouchaud2020much} hold in a more general context, in particular the emergence of inequalities when productivity is increased when agents saturate their consumption. 

The link between prices, income, supply, and demand has been long studied in the field of economics \cite{gale1955law,schultz1935interrelations}. Unfortunately, the two-good toy model explored in \cite{bouchaud2020much} does not generalize well to a multi-good situation. In the present work, we aim to expand this model to include a larger array of products and treating each individual as an independent agent acting from personal preferences rather than a set of preordained rules, resulting in a complex disordered system. Still, we find that as the ``productivity'' of the economy increases, the systems undergoes a phase transition between a low-inequality economy and a high-inequality economy. The transition occurs due to a rather subtle collective instability, and not due to the existence of particularly attractive goods (or skilled agents). We numerically establish a variety of interesting phenomena that take place in our highly stylized economy. We justify some of our results using analytical calculations.

The paper is organised as follows -- in sec.\ref{sec:ABMdef}, we establish the behaviour of individual agents in terms that make sense on a physical level of a society, and examine behaviour such a system exhibits numerically in sec.\ref{sec:obs}. To understand the mathematical reasons behind the phase transitions we observe, we then simplify the system as a set of continuous differential equations in sec.\ref{sec:contdef} that maintains the features previously observed while being analytically tractable. In sec.\ref{sec:rmtsols} we perform a linear stability analysis of these equations using Random Matrix Theory methods. Finally, in secs.\ref{sec:taxes} and \ref{sec:altdyn} we discuss possible solutions to economic inequalities arising within the model before concluding.

\section{An Agent-Based Model}

\subsection{\label{sec:ABMdef}Trades and Prices}

For the basic agent-based model, we imagine a system consisting of $N$ agents, who are all producers of one good each, and consumers of any goods they can buy, with some preferences defined below and an upper limit for each good, beyond which there is no further increase of satisfaction. This setting can be an approximation for people living in a historic trades-based society, some sort of barter system between farmers, or a combination of firms and individuals in a more modern sense. 

Each agent $i=1, \dots, N$ produces $Y_i:=y_i N$ units of good $i$, and attempts to sell it at a price $p_i$. (In the following we will set $y_i = y \;\forall \;i$.) Agent $i$'s preference for (or utility gained from) the good produced by agent $j$ is an independent random variable $J_{ij} \geq 0$, with a given distribution, identical for all agents. In the following, we will often consider an exponential distribution $\rho(J)=e^{-J}$, although most of our results are qualitatively the same for other distributions.  

\begin{figure}
    \centering
    \includegraphics[width=0.5\linewidth]{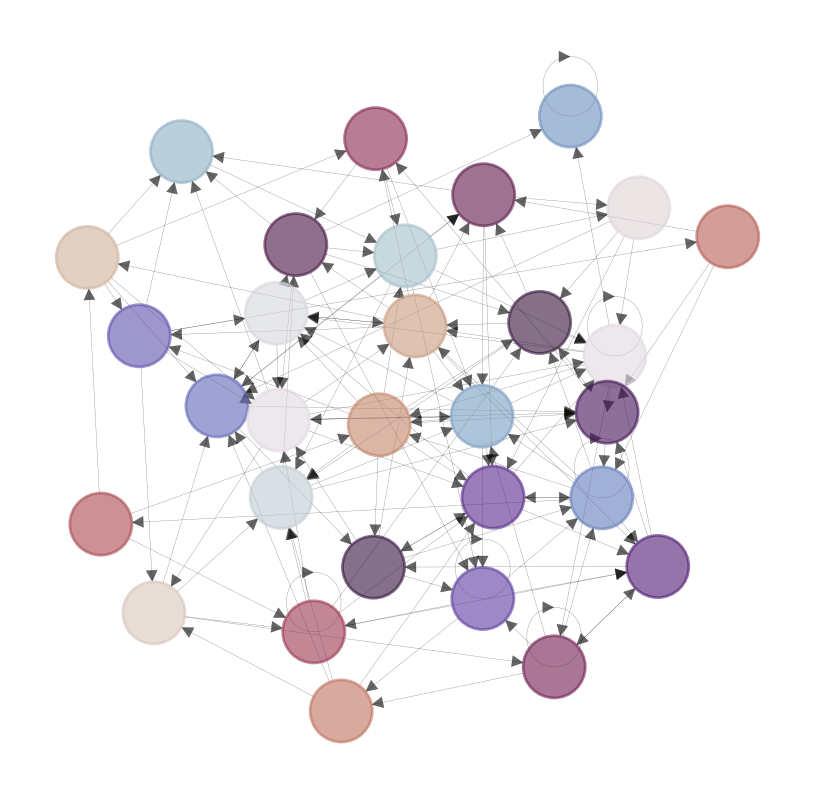}
    \caption{An example of a small system following the rules of our model at a particular point of time, where the lines show money flowing from one agent to another if they bought a good from them.}
    \label{fig:networkplot}
\end{figure}
Let the network have an adjacency matrix $C_{ij}$. The system starts off with everyone having the same amount of wealth $w_i(t=0)=W$ and selling their goods at a random price $p_i(t=0)=\left[\mathcal{N}(1,0.2)\right]^+$ centred around $1$. We assume there is a quantum of goods $\gamma$ which is the minimum amount of a product that an agent can buy,  and there is a maximum $\Gamma$ at which the utility saturates. For simplicity, we take $\gamma=\Gamma=1$, which means that each agent can either buy one unit of good or abstain. In other words, one cannot buy a fraction of an apple and agents do not care for more than an apple a day. However, all the results below hold for general values of $\Gamma$ and $\gamma$, provided goods remain indivisible.

To maximise their utilities given a limited budget on day $t$, agents buy products with the highest utility per unit price $J_{ij}/p_j$ first, and proceed to buy goods with lower ``value for money'' after saturating themselves on more preferable goods. They continue this process till they have no money left, or till they have bought everything they wanted to buy. They use any money they make from people buying their goods only on the next day, i.e. at $t+1$, i.e. money cannot be borrowed or lent. Thus the change in wealth $\Delta w_i$ for each agent on a particular day is given by
\begin{align}
    \Delta w_i=p_i\sum_jT_{ji}-\sum_jT_{ij} p_j\label{eq:Dworig}
\end{align}
where the terms $T_{ij} \in \{0,1\}$ signify the units of product $j$ bought by agent $i$ on a particular day. Note that $T_{ij}$ a priori depends on the wealth $w_i$ of agent $i$, the price $p_j$ of good $j$ and the propensity for agent $i$ to buy product $j$, $J_{ij} \times C_{ij}$. 

To optimise sales, agents reduce their price if they sell less than what they produce, and increase it if there was a high demand even if they ran out of stock. We posit that log-prices evolve linearly as a function of the relative supply-demand imbalance:
\begin{align}
    \log\left(\frac{p_i+ \Delta p_i}{p_i}\right)= \frac{r_p}{Ny_i} \sum_j (D_{ji}-y_i) \label{eq:Dporig}
\end{align}
where we make a distinction between the ex-ante demand $D_{ij}$ and the actual purchases $T_{ij}$ to allow the demand to possibly exceed the realised sales. The coefficient $r_p$ measures how fast prices adjust to supply-demand imbalances. 

It is easy to see that the equilibrium conditions of our economy are
\begin{align}
    \begin{aligned}
        \sum_jT_{ji}&=y_i N\\
        \sum_jT_{ij}p_j&=\sum_jT_{ji}p_i=y_i Np_i.
    \end{aligned}\label{eq:equil_eqs}
\end{align}
Note that these equations are invariant under a uniform rescaling of prices, i.e. our model, like the true economy, must be inflation invariant. Considering $y_i \equiv y$ and summing over a large number of agents, this gives that the probability of the binomial variable $T$ being $1$ is $y$, i.e. there is on average a probability $y$ that any agent buys the product sold by any other agent. When $y \leq 1$, agents should thus be able to coordinate in such a way that markets clear, since this would earn each agent enough income to buy $yN$ units of goods among $N$. When $y > 1$, on the other hand, the constraint that the amount of each good cannot be more than $\Gamma = 1$ necessarily leads to overproduction and deflation. However, as will be clear below, the price elasticity of demand is reduced  as $y$ increases towards $1$. This effect is responsible for an incipient instability of the economic equilibrium of the system at a value $y_c < 1$, beyond which coordination fails and wealth inequalities appear.

\subsection{Observations\label{sec:obs}}

\begin{figure}\hspace{-0.8cm}
    \includegraphics[width=\linewidth]{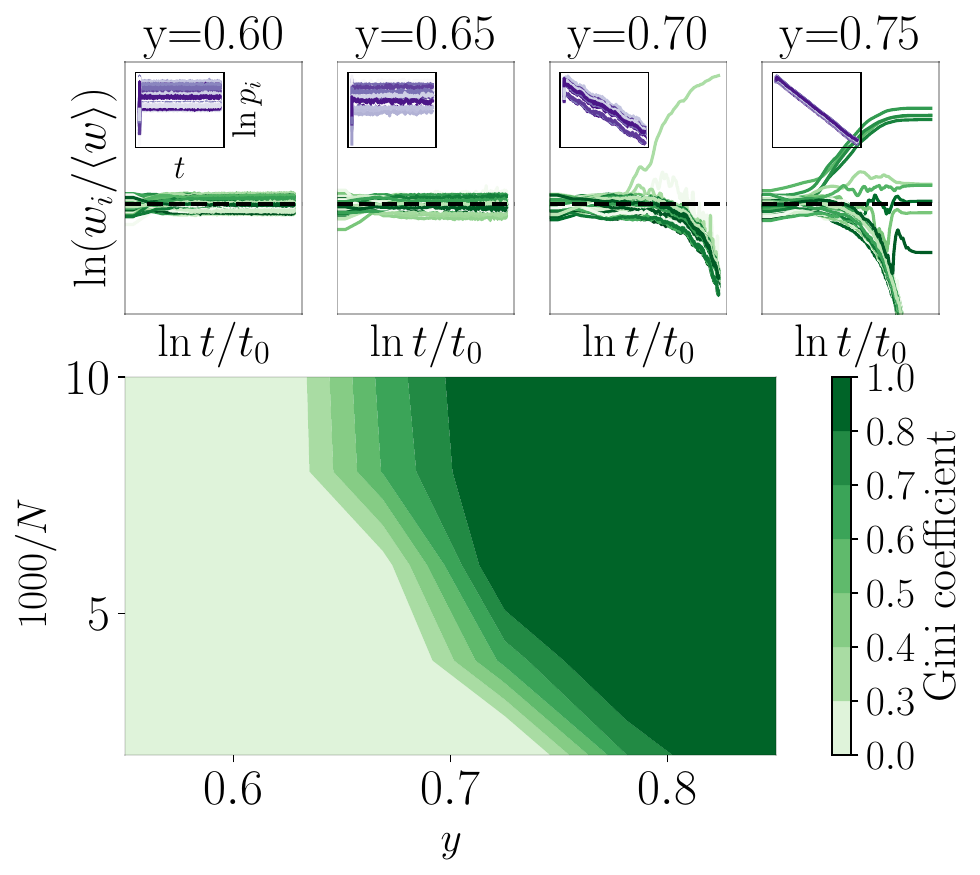}
    \caption{The dynamics of wealth and prices in this system exhibit a phase transition as we increase the amount of goods produced, leading to price deflation and the concentration of wealth in the hands of a few individuals. {\bf Top}: individual wealth trajectories for different values of $y$. In the insets in purple, we show the corresponding trajectories of individual goods' log prices with (linear) time. For $y > y_c$, deflation sets in, and creates wealth inequalities. {\bf Bottom}: Using the Gini coefficient as a phase parameter and a measure of inequality, we see in the phase diagram that for smaller networks this transition to large inequalities happens at lower values of $y$ than for larger systems. }
    \label{fig:origmodel_timeseries_phase}
\end{figure}
On running simulations for this rather simplistic model for what can be considered a primitive barter style economy, we find that for low values of productivity $y$, everyone maintains a similar amount of wealth to what they start with. Most surprisingly, however, at a certain critical value of the productivity $y_c < 1$ (\cref{fig:origmodel_timeseries_phase}), there is a transition to a society where the wealth gets concentrated in the hands of a few people (the ``oligarchy'' \cite{boghosian2019inequality}), and the rest have nothing left, with permanent under-consumption and price deflation, see \cref{fig:origmodel_timeseries_phase}. There is a variety of indicators that can be used to quantify inequalities. We shall be using prominently the Gini coefficient as our phase parameter, as it adequately captures the kind of inequality we are dealing with. In \cref{fig:origmodel_timeseries_phase}, we see that the transition in the Gini coefficient is abrupt, suggesting a first-order phase transition in the system. We also see that besides the productivity $y$, the system size $N$ also plays a role in determining when the transition occurs, and smaller systems lead to faster transitions. The role of $N$ will be clarified by our analytical calculations below. 
 \begin{figure}
     \centering\includegraphics[width=\linewidth]{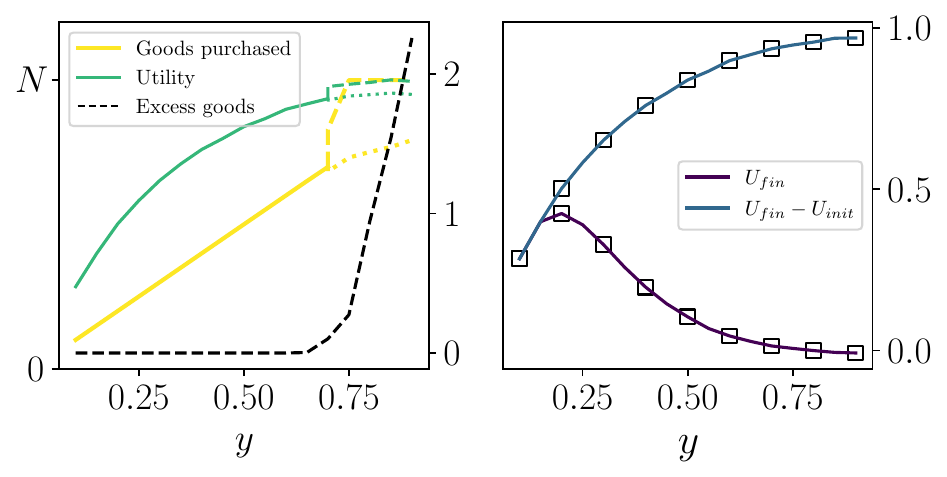}
     \caption{For a small system of $N=100$ agents, we look at how goods exchange dynamics changes at transition. {\bf Left}: The dashed black line, right scale, represents the average surplus production of agents, which seems to monotonically increase from zero as soon as the phase transition occurs. The coloured lines representing utilities and number of goods purchased split into dashed and dotted lines to show the emergent class division after the phase transition. In the unequal phase, we see that upper class agents can buy every product in the market, while the lower class buys on average 20 fewer goods. On the other hand, the difference between utilities of the two classes is marginal. The increase in unsold goods is equivalent to the inflation (or rather deflation) rate of prices visible in \cref{fig:origmodel_timeseries_phase}. {\bf Right}: The top curve shows the average utility of the final state, and the bottom curve shows the difference of utility between the final state and the initial state, both as a function of $y$. We can see that while the total utility grows as a function of $y$ as people are able to buy more products, the system shows a much higher improvement of utility with time for lower values of $y$.}
     \label{fig:Upurchsurp}
 \end{figure}

Aside from the rise of inequalities, the phase transition signals many other changes in the microscopic details of the system. In \cref{fig:Upurchsurp} (left), we see that along with the Gini, there is a rise in unsold goods as shown by the dashed black line. The amount of goods purchased by people also starts showing a marked difference between the upper and lower classes at this point. If we choose to define individual utilities as the sum of the utilities of the goods purchased $U=\sum_jT_{ij}J_{ij}$, we see that there appears to be only little difference between the two classes. However, considering the vast difference in the savings of the two classes, having any other avenues for spending or investing wealth would reveal a much larger difference in utilities. This means that the introduction of any good external to the system, the increase of the limit $\Gamma$ on each goods purchased (set to unity above), or even simply considering the security felt by the owner of a large bank account, would produce high disparity between upper and lower classes. This also does not consider utility functions that include more than just a sum of individual purchases, for example inserting a term that reduces the global utility when there are large inequalities, if products go unsold, or for large deviations of production from the ``natural'' initial state. In other words, within the strict setting of the model where it is impossible to spend more than the cost of the $N$ different goods, the accumulation of wealth is pointless and has no strong effect on the rest of the society. But clearly, the appearance of unspent wealth opens the door to a capitalistic society. 
 
To get a better understanding of what determines these emergent inequalities, consider the matrix $T_{ij}^{eq}=\Theta(J_{ij}+\log y)$ for exponentially distributed $J_{ij}$. This is a mean-field approximation for which products are being bought by whom in equilibrium, which become binomial variables with probability $y$ of being $1$ satisfying \cref{eq:varppred}, \cref{eq:varSpred} as needed. It will be useful to relax our definition of the Heaviside function, for example by using the Fermi-Dirac function 
\begin{align}
    \Theta_\beta(E):=\frac{1}{1+\exp(-\beta E)},\label{eq:fermidirac}
\end{align}
which reduces to a $0,1$ function when $\beta \to \infty$. We can now calculate the associated smooth delta-function as
\begin{align}
    \delta_\beta(E):= \frac{\partial \Theta_\beta(E)}{\partial E} = \frac{\beta\exp(-\beta E)}{(1+\exp(-\beta E))^2}\label{eq:deltafunc}
\end{align} 
The idea is that the final state of our primitive economy should be to a large extent entirely determined by the structure of the preference matrix  $J_{ij}$ and hence by its dependants $T_{ij}^{eq}$ or $\partial T_{ij}^{eq}$, given by the above smooth functions applied on $E_{ij}=J_{ij}+\log y$. To this end, we calculate various centrality measures using these matrices, like page-rank, eigenvector centrality, in-degree, and closeness, among others (see e.g. \cite{newman2018networks} for definitions). We find that the in-degree -- calculated as a sum of the weights of each in-edge given by the values in the matrix corresponding to an agent -- correlates best with final values of wealth in the system. Page-rank, closeness, and eigenvector centralities also correlate very well with final wealth, while many other measures of centrality seem to have no relation to the system observables. 

In \cref{fig:centcorrs} (left), we see that the mean-field $T_{ij}^{eq}$ explains people's position in wealth ranking better than the $J_{ij}$'s themselves, but our ability to guess individual wealth using either of these matrices drops significantly beyond the transition point $y=y_c$. In fact, while the correlation with rank wealth remains significant, the drop in correlation with actual wealth shows that there is not much reason behind the enormous amount of wealth certain individuals have. 
In other words, even if their are ``objective'' reasons for some agents to gather more wealth than others, the breakdown of coordination taking place beyond $y > y_c$ leads to price deflation, which in turn amplifies considerably wealth inequalities. In \cref{fig:centcorrs} (right) we show the relation between the ``survival probability'' -- defined as the probability of being in the upper class -- as a function of the in-degree of each agent. While our system is technically fully connected, the highly non-linear nature of the Heaviside function is in this way reproducing the effects of graphs with non-uniform degree distributions \cite{aguirre2024heterogeneous,park2024incorporating,poley2024interaction}, which can be further investigated on different connectivity structures. 

\subsection{A collective phase transition\label{sec:collective}}

\begin{figure}
    \centering
    \includegraphics[width=\linewidth]{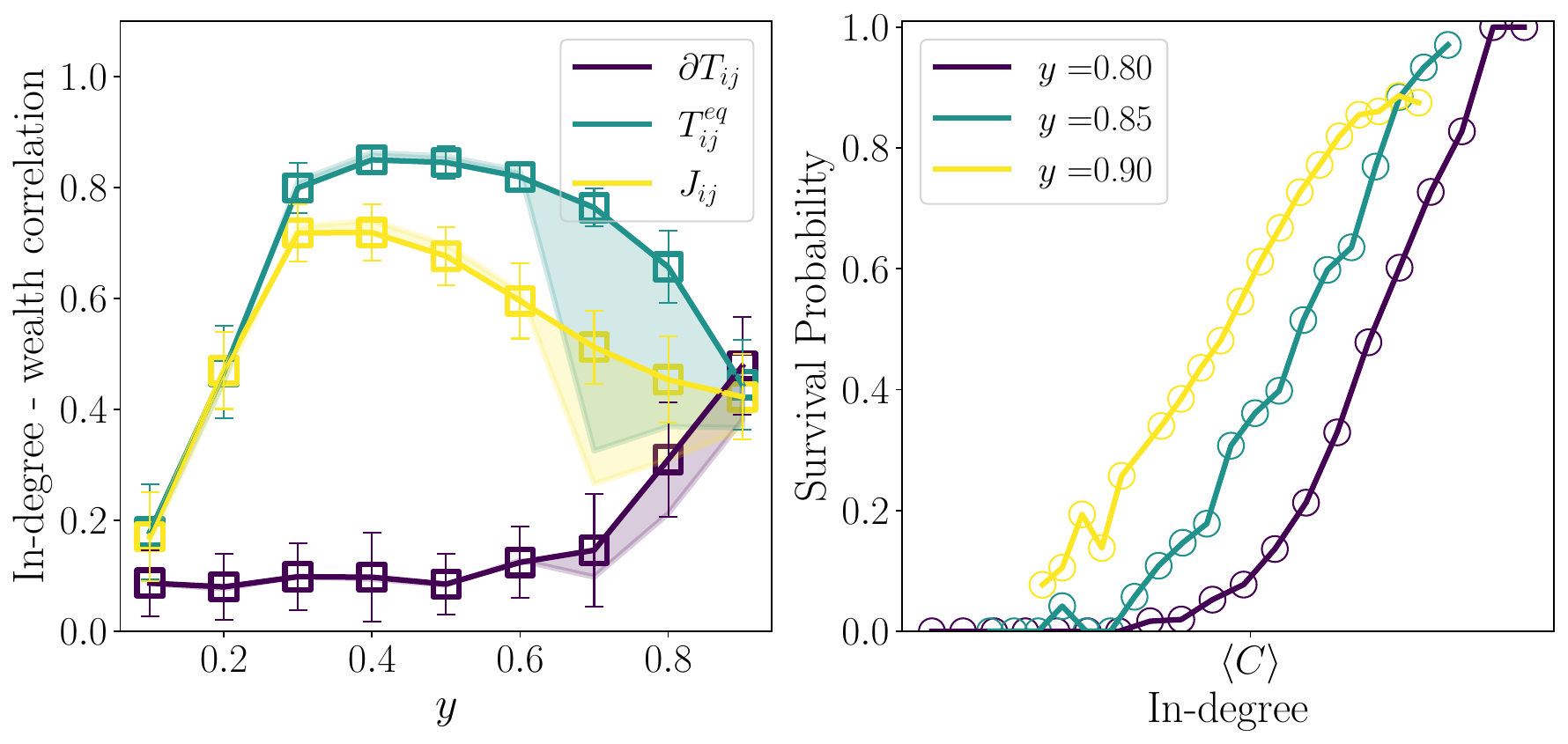}
    \caption{{\bf Left}: The solid lines represent the correlation of ``in-degree" of nodes for various matrices with rank of the wealth owned by agents, with error-bars representing their spread over 20 simulations. Values are calculated at the end of $T=2000$ time steps for a system of $N=100$ people. The shaded regions represent the difference between the correlation observed for ranked wealth with that observed for wealth itself. {\bf Right}: For different values of production $y$, we see the dependence of the ``survival'' probability on the in-degree. The survival probability is defined as the probability of an agent to belong in the upper class of the system post transition.}
    \label{fig:centcorrs}
\end{figure}

As we will see below, the phase transition in our system is truly collective and related to non-local, multi-step trading loops -- i.e. I buy from you and you buy from her and hence she can buy from me.  
Indeed, what stands out most in \cref{fig:centcorrs} is that for systems after transition, rank wealth correlation with the measures of the derivative matrix $\partial T_{ij}^{eq}$ increases. As we are in this case looking particularly only at the sum of values along columns of the matrix itself, this suggests that the richest individuals are not just those who have many clients, but also clients that are right at the edge in their decision making of whether or not to buy. The fact that $T_{ij}^{eq}$ explains wealth differences better than $J_{ij}$ suggests that the non-divisibility of goods creates inequalities in the model. 

In this regard, assume instead a world where goods are infinitely divisible. Using a standard logarithmic utility function, one finds that the consumption of good $j$ by agent $i$ is given by $T_{ij} = \mu_i J_{ij}/p_j$ where $\mu_i$ is such that the budget of agent $i$ is balanced. Adding market clearing, the equilibrium equations of the model then read
\begin{align}
    \begin{aligned}
        \sum_j J_{ji} \mu_j -  \mu_i \sum_j J_{ij} &= 0\\
         \mu_i \sum_j J_{ij} &= Ny p_i
    \end{aligned}\label{eq:classical_eq}
\end{align}
These equations always have positive solutions for $\mu_i$ and $p_i$, with $p_i$ independent of $y$ and $\mu_i(y) = y \mu_i(1)$. In this context, nothing much happens as $y$ is increased (see also Appendix \ref{sec:classical} where we discuss the dynamical stability of this equilibrium). 

\subsection{Adjustment of production capacities}

It might be perceived as strange that in the $y > y_c$ phase our agents continue producing $yN$ products every time step (or continue treating $yN$ as the standard of number of products that should have been sold) despite facing major losses and scarce demand, as evidenced in \cref{fig:Upurchsurp} (left). Imagining our agents to have a say in how much they produce, we consider an additional scenario where they adjust their individual $y_i$'s based on how much they sell at speed $r_y$. This  gives us an extra equation for the dynamics of $y_i$:
    \begin{align} 
    \begin{aligned}
    \log\left(\frac{y_i+ \Delta y_i}{y_i}\right)&= \frac{r_y}{Ny_i} \sum_j (D_{ji}-y_i) \\&\equiv \frac{r_y}{r_p} \log\left(\frac{p_i+ \Delta p_i}{p_i}\right) \end{aligned}\label{eq:DYorig}
    \end{align}

\begin{figure}
    \centering
    \includegraphics[width=\linewidth]{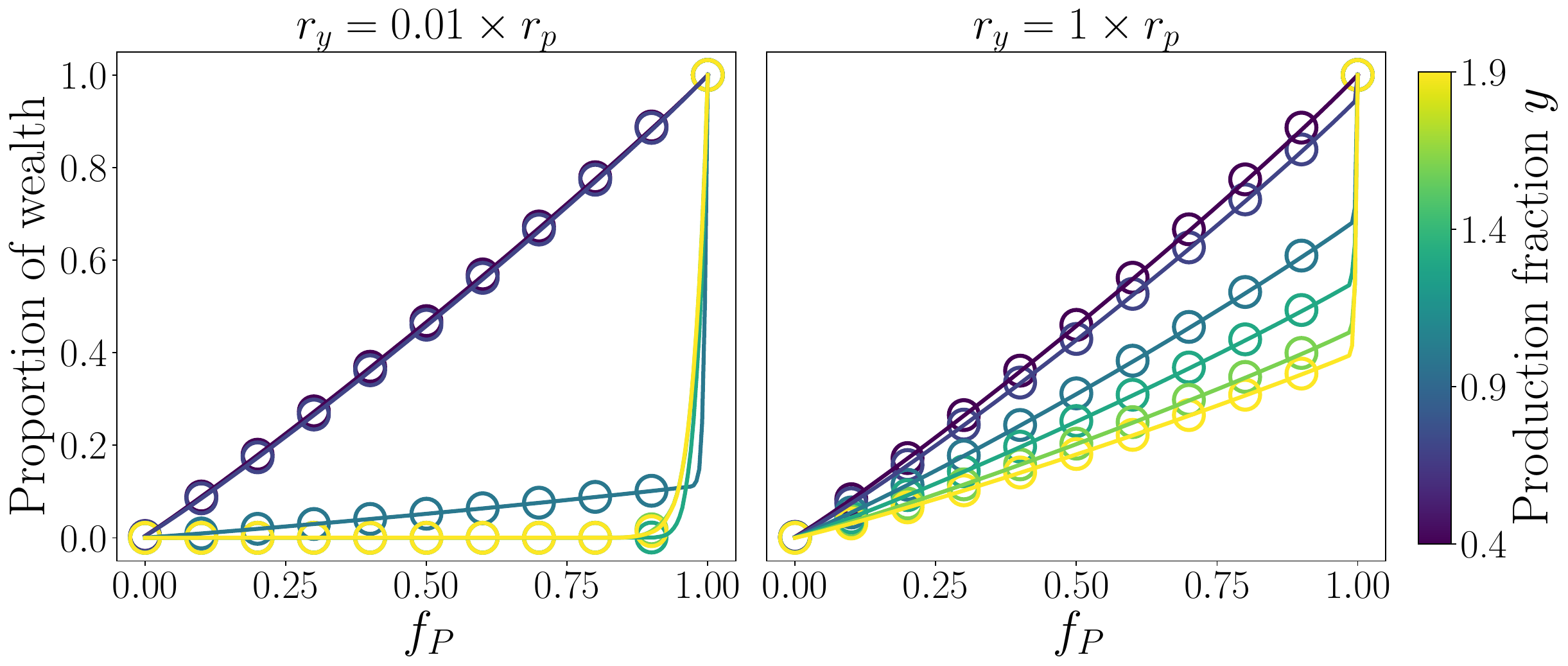}   
    \caption{Lorenz curves, i.e. plots of the fraction of wealth owned by a certain quantile of the population $f_P$, calculated after a certain fixed number of time steps ($T=5000$) for simulation runs with different values of the initial production benchmark $y_0$ and  $r_y/r_p=0.01$ (left) and $r_y/r_p=1$ (right), and for $N=200$. We see that beyond a certain value of $y$, a steep discontinuity appears for $f_P \sim 1$, revealing the appearance of a rich class which owns most of the wealth. }
    \label{fig:Lorenzcurves}
\end{figure}

This system still has the same equilibrium as before, but the erstwhile instabilities only result in reducing the $y$'s till the system is just stable again. This is an example of ``self-organised criticality'' \cite{bak2013nature}: starting the system off at $y_i = y > y_c$ eventually equilibrates around the critical point $y_i\approx y_c$. 
In \cref{fig:Lorenzcurves}, we plot the distribution of wealth in the system via the Lorenz curves (proportion of total wealth owned by the poorest $f$ fraction of society), for different values of $y$ and $r_y/r_p$. We see that inequalities in the case of adjustable production is less drastic but keeps on increasing with $y$.
\begin{figure}
    \centering
    \includegraphics[width=\linewidth]{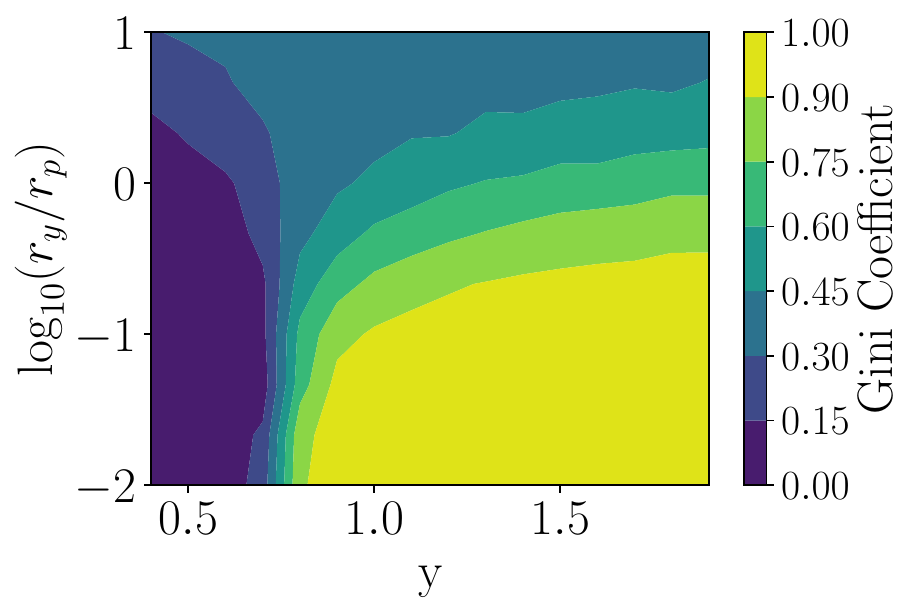}
    \caption{Allowing agents to adjust their production leads the system to always find a steady state, where the level of inequalities depends on initial values of $y$ and the rates at which prices and productions adjust. For low production change rates we recover the discrete jump in Gini at $y_c$ showcased in\cref{fig:origmodel_timeseries_phase}. For high production adjustment rates, the Gini asymptotically approaches a value between 0.3 and 0.4 for any initial value of production, close to the critical value obtained for $r_y=0$.  Here $N=200$ and $r_p=0.01$.}
    \label{fig:prodpricerates}
\end{figure}

For higher and higher values of production change rates $r_y$, we find that the system stabilises to a default state of inequalities with a Gini coefficient of around 0.4, as seen in \cref{fig:prodpricerates}. This value corresponds to the critical Gini coefficient for $y=y_c$ when $r_y=0$, i.e. the system self-organises at criticality. The corresponding Lorenz curve describes a society where less than 1\% of people own almost 40\% of the total wealth. Interestingly, most recent estimates of global wealth inequalities contain similar observations \cite{riddell2024inequality}, though this is most likely a coincidence, as our model is not calibrated to the real world. In fact, the Lorenz curve and the Gini coefficient depend in our model on the rate of change of prices $r_p$ as well.

\section{Analytical results}

While this suffices as the entire description of the system as far as numerical simulations are concerned, we realise that it is difficult to write an analytical expression for the matrices $T$ and $D$. To deal with this problem, we introduce a dummy variable $S_i$, called the threshold, signifying the minimum acceptable \textit{value for money} of products for agent $i$ to buy something. We will work with a fully connected network of individuals, letting us write the demand as
\begin{align}
    D_{ij}=\Theta(J_{ij}/p_j-S_i)=\Theta(J_{ij}-p_jS_i)\label{eq:defthreshold}
\end{align}
Essentially, agents calculate their thresholds on each day such that they will not overshoot their budgets. Again, $T_{ij}$ is somewhat difficult to define giving limited stocks for all goods, and would depend on how much stock of a product is left when agent $i$ arrives to buy product $j$. We find that it does not really matter even if we let agents oversell slightly, so for simplicity we assume they have an infinite stock, $yN$ being how much they prefer to be producing and not necessarily a hard limit. This lets us set $T=D$. 

\subsection{Mean-field Equilibrium}

Since all agents are statistically identical, we expect that for large $N$ they will all sell at roughly the same prices and have the same consumption threshold. Hence we make the following {\it ansatz}: 
\begin{align}
\begin{aligned}
    p_i&=p^\star \left(1+\frac{\epsilon_i}{\sqrt{N}}\right) + O\left(\frac1N\right),\\ S_i&=S^\star \left(1+\frac{\zeta_i}{\sqrt{N}}\right)+ O\left(\frac1N\right),
\end{aligned}
\end{align}
where $p^\star $ and $S^\star $ are the average values of the prices and thresholds at equilibrium. 
Doing a first order expansion on one element of $T$ with respect to $N^{-1/2}$ then gives
\begin{align}
    \begin{aligned}
T_{ji}&=\Theta(J_{ji}-p^\star S^\star (1+\frac{\epsilon_i}{\sqrt{N}}+\frac{\zeta_j}{\sqrt{N}}))\\
&=\Theta(J_{ji}-p^\star S^\star )+p^\star S^\star \delta(J_{ji}-p^\star S^\star )\left(\frac{\epsilon_i}{\sqrt{N}}+\frac{\zeta_j}{\sqrt{N}}\right)\end{aligned}\end{align}
Since the overall equilibrium price level is arbitrary we can always set $p^\star =1$. Thus the demand for the good $i$ is \begin{align}\begin{aligned}
\sum_jT_{ji}&\approx NP_>^\star +\sqrt{NP_>^\star (1-P_>^\star )}\eta_i-\rho^\star S^\star  \sqrt{N}\epsilon_i
    \end{aligned}
\end{align}
where $\eta_i$ is a Gaussian white noise, $P_>^\star =\mathbb{P}(J>S^\star )$, $\rho^\star =\rho(J=S^\star )$. We get the above expression via the Central Limit Theorem after treating the Heaviside function as a binomial variable with $P_>^\star $ probability of being $1$. As this demand must be equal to $Ny$ in equilibrium, we should have
\begin{align}
    P_>^\star =y \qquad {\rm and} \qquad\epsilon_i=\frac{\sqrt{P_>^\star (1-P_>^\star)}}{S^\star \rho^\star }\eta_i
\end{align}which tells us that at equilibrium the prices have a spread of
\begin{align}
    \sigma_p^2=\frac{y(1-y)}{N(S^\star \rho^\star )^2}\label{eq:varppred}
\end{align} For the thresholds, we use the second equation in \cref{eq:equil_eqs} letting us deduce that
\begin{align}
    \begin{aligned}
        \sigma_S^2=\frac{y(1-y)}{N \rho^{\star 2}}\left[1+\frac{y^2}{(S^\star \rho^\star )^2}\right]
    \end{aligned}\label{eq:varSpred}
\end{align}
\begin{figure}
    \centering
    \includegraphics[width=0.6\linewidth]{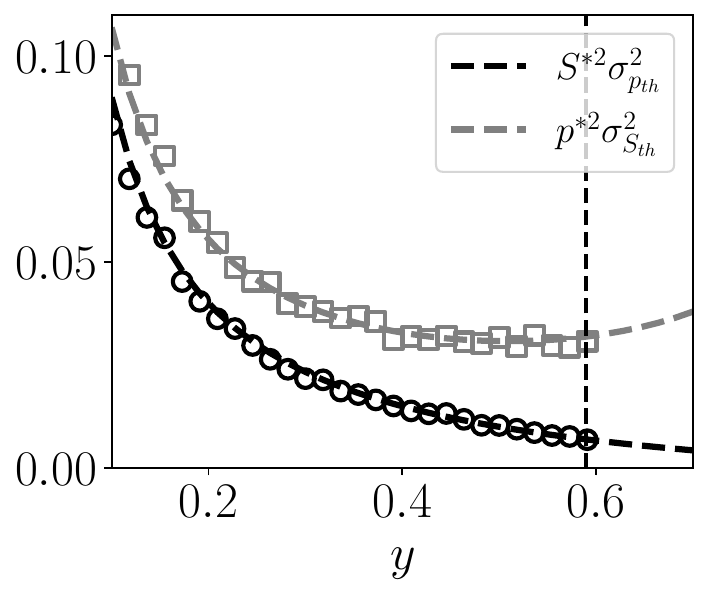}
    \caption{The observed variances in thresholds and prices in simulations compared with theoretical predictions (dashed lines) from \cref{eq:varppred,eq:varSpred}, for $N=100$ and $\rho(J)=e^{-J}$. We observe strong deviations from theoretical predictions at the value of $y$ at which the phase transition occurs. The corresponding data points after the transition are off-scale and not shown.}
    \label{fig:var_pS}
\end{figure}
We see in \cref{fig:var_pS} that our predictions have some merit, as the variances of prices and thresholds observed in simulations follow closely \cref{eq:varppred,eq:varSpred}, but fail to hold beyond a certain value $y = y_c$ at which the transition occurs. As these equilibrium equations contain no indications of instabilities or inconsistencies for any value of $y<1$, we must use other methods to predict when and understand why such a transition occurs.

\subsection{Continuous time approximation\label{sec:contdef}}


We started our work from a physical description of a society where people produce and exchange goods for wealth in sequence of importance, limited by amount of wealth available. We have then considered an approximation where the threshold $S$ is a proxy for the wealth $w$, calculated by considering the affordability of items thus as a function of the wealth. In the first description, if a product costs more than an agent can afford, they try to buy the next best product if they can afford it. In the second, they stop purchasing anything once they hit the first such item. These two processes are almost equivalent, as the instances in which affordable products are ranked after expensive ones are limited, leading to negligible quantitative differences in our findings between agent based simulations of the two. As a final simplification, we introduce continuous time, and approximate the thresholds to be defined by the wealth in a way that the probability of buying a good depends on what fraction of products an individual might be able to afford given their wealth. We thus write two evolution equations, one for price and one for wealth, which reflect the same logic as \cref{eq:Dporig,eq:Dworig}: 
\begin{align}
    \begin{aligned}
     \frac{d p_i}{dt} &= \frac{\kappa p_i}{Ny}  (\sum_jT_{ji} - Ny) \\
      \frac{d b_i}{dt} &= \frac{\kappa}{N} \left(p_i\sum_jT_{ji}-\sum_jT_{ij}p_j\right)
    \end{aligned}\label{eq:cont_pw}
\end{align}
where $b_i:=w_i/N$ can be considered the ``budget'' of an agent for each good. We define $T_{ij}$ as 
\begin{align}
  T_{ij} = \Theta\left(J_{ij}-\frac{p_j}{\overline{p}}P_>^{(-1)}\left(\frac{b_i}{\bar{p}}\right)\right), \label{eq:newdefTijgen}
\end{align}
which comes from estimating the threshold in terms of prices and budgets in equilibrium, as to satisfy on average \cref{eq:equil_eqs}. For exponentially distributed utilities one gets 
\begin{align}
    T_{ij}=\Theta\left(J_{ij}-\frac{p_j}{\overline{p}}\log\frac{\overline{p}}{b_i}\right)\label{eq:newdefTij}
\end{align}

These equations have the same equilibrium as suggested in \cref{eq:equil_eqs}. To be able to perform a linear perturbation analysis on this, we take the Heaviside to be given by the Fermi-Dirac function as in \cref{eq:fermidirac}. To match the original model quantitatively, we identify the rate of change of prices in the continuous and the ABM versions  given a small perturbation, which gives
\begin{align}
\begin{aligned}
        \frac{\delta_\beta(0)\kappa {\rm d}t}{Ny}&=\frac{r_p}{Ny}
        \implies\beta=\frac{4r_p}{\kappa {\rm d}t}
\end{aligned}
\end{align} where $\delta_\beta$ is the function in \cref{eq:deltafunc} and causes the discreteness in the continuous simulations, while in the original discrete model the price change is just one times the rate $r_p/N$. We see with the chosen expressions that the wealth change then matches up for the continuous and the agent based model. 

The addition of an adjustment rate for production would give us an additional equation
\begin{align}
    \begin{aligned}
              \frac{d y_i}{dt} &= \frac{\kappa'}{N}  (\sum_jT_{ji} - Ny_i) 
    \end{aligned}\label{eq:Yadjusteqs}
\end{align} alongside \cref{eq:cont_pw} where we replace $y$ with $y_i$, and assuming that $\kappa'$ ($\propto r_y$) in the same way as $\kappa$ ($\propto r_p$), i.e. $\kappa'=\kappa r_y/r_p$.


\subsection{Linear Perturbation Analysis\label{sec:rmtsols}}
 We can now linearise the dynamics \cref{eq:cont_pw,eq:Yadjusteqs} around equilibrium to give a block structure stability matrix 
\begin{align}
\begin{aligned}
        \begin{pmatrix}    \Dot{\Vec{\delta p}}\\\Dot{\Vec{\delta b}}
    \end{pmatrix}&=\begin{pmatrix}
    \mathbb{A} &\mathbb{B} \\\mathbb{C} & \mathbb{D}
    \end{pmatrix}\begin{pmatrix}
    \Vec{\delta p}\\\Vec{\delta b}
    \end{pmatrix}
\end{aligned}\label{eq:veceqdef}
\end{align} The elements of the blocks in \cref{eq:veceqdef} are given in Appendix \ref{sec:blockterms}. It is easy to see that the constant matrix composed of the blocks in \cref{eq:constmatgen} has one eigenvalue equal to $\lambda_0=0$, and that it corresponds to a mode given by the eigenvector $v_0=\begin{pmatrix}
    \begin{pmatrix}
        1 & 1 & \dots & 1
    \end{pmatrix}&\begin{pmatrix}
        y & y & \dots & y
    \end{pmatrix}
\end{pmatrix}^T$. This mode turns out to be related to the invariance of our economy under a rescaling of all prices $p$, provided wealth is similarly scaled such that $b/p$ remains a constant. Aside from this mode, there are $N-1$ eigenvalues at  $\lambda_p=\kappa\ln y$ (for an exponential distribution of $J$), and $N$ at $\lambda_b=-\kappa$. We can conclude that the eigenvalues of the actual matrix must be distributed around these values, which we notice are negative as long as $y<1$, but $\lambda_p$ goes to zero proportionally to $-\kappa (1-y)$ when $y \rightarrow 1$. As shown in Appendix \ref{sec:blockterms}, this is actually a general result, for arbitrary distributions of $J$. Intuitively, this is associated to the fact that the dependence of demand on prices becomes very weak when $y \to 1$, in such a way that the stabilising feedback loop between prices and demand stops operating. In Appendix \ref{sec:classical}, we show that this is not the case when good quantities are continuous and demand is set by classical utility maximisation. 

Another interesting fact is that the leading mean $\lambda_p$ corresponds to eigenvectors that satisfy
\begin{align}
    \Vec{\delta b}=y\frac{\ln y+1}{1-\kappa\ln y}\Vec{\delta p}, \qquad  \sum_j\delta p_j=0
\end{align}
Interestingly, these modes correspond to price and wealth conserving fluctuations. 
\begin{figure}
    \centering
    \includegraphics[width=\linewidth]{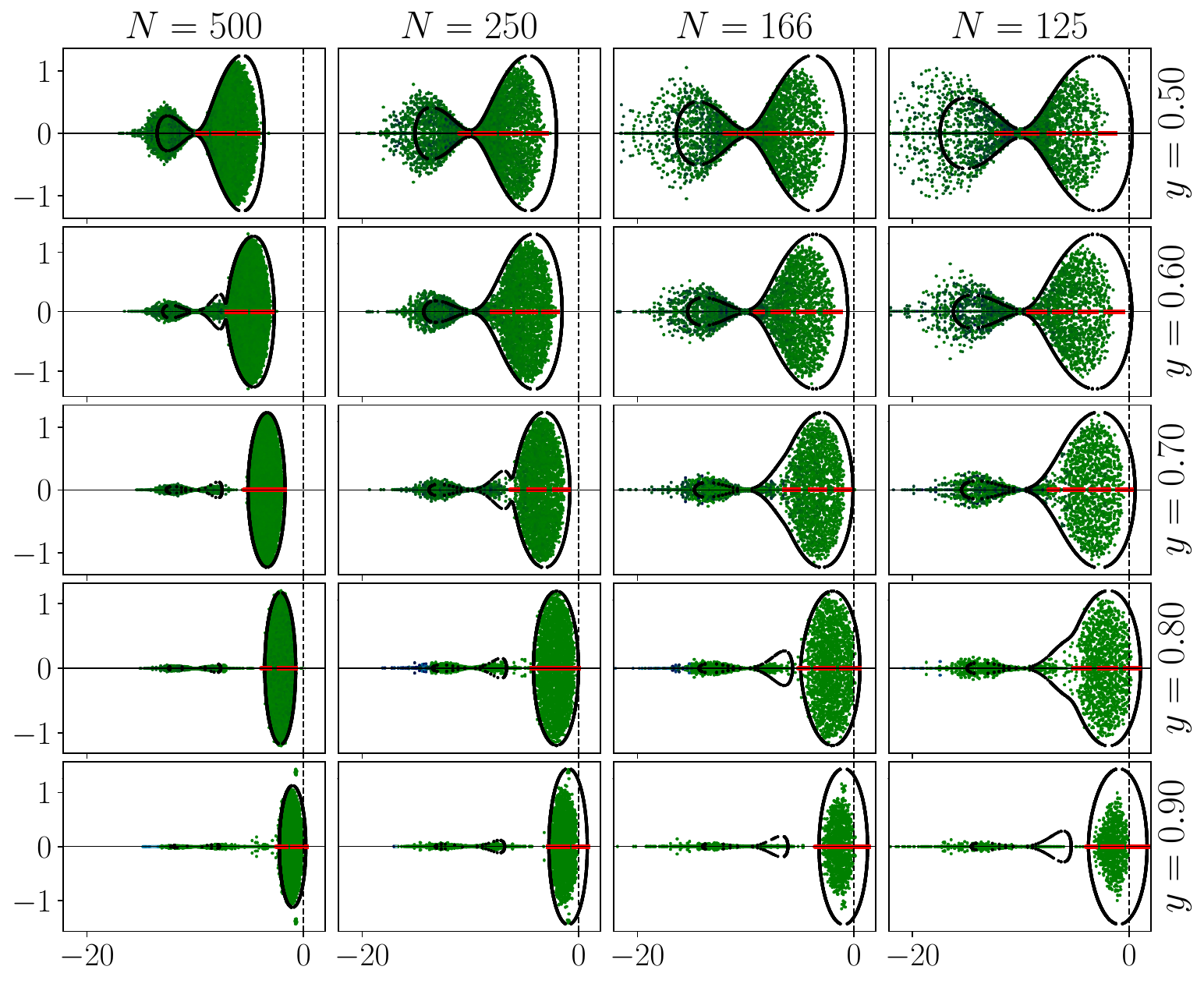}
    \includegraphics[width=\linewidth]{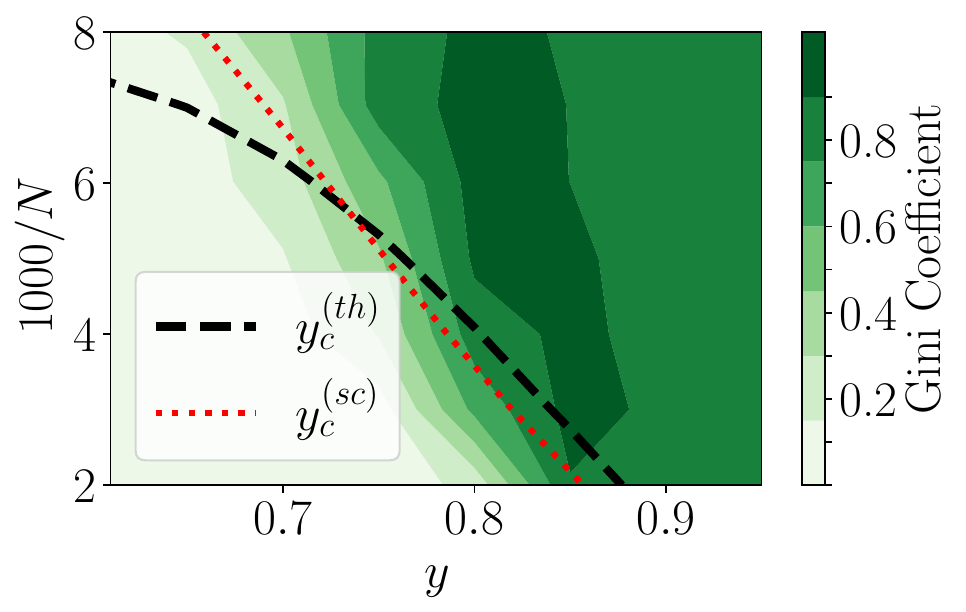}
    \caption{{\bf Top}: A direct comparison between the spectra of the linear stability matrices of the model with predictions for the boundaries of the spectra from \cite{patil2023stability}. The match is better for larger systems, as our assumptions on the uniformity of equilibrium variables has a smaller error there. {\bf Bottom}: Theoretical predictions for where the spectrum crosses the imaginary axis on the phase diagram using the Gini-coefficient as the phase parameter. In black we have the prediction $y_c^{(th)}$ coming from the exact positions of the boundary of the spectrum calculated using the correlations in Appendix \ref{sec:blockterms}. In red, we have the scaling argument prediction from \cref{eq:scalingargs}, which is also used to construct the rough bounds $y_c^{(sc)}$ in red on the left bulk of the spectra, with length $\sqrt{{\rm Tr}[\Psi_0]/N}$, with $\Psi_0$ being the covariance matrix of a block within the rearranged stability matrix \cref{eq:blockelements}. }
    \label{fig:spectravsmodel}
\end{figure}
Knowing the correlations between the elements of the stability matrix at hand, we use random matrix methods developed  in \cite{patil2023stability} to find the boundary of the eigenvalue spectra of the system and determine the exact point at which this transition occurs. This method uses the correlations \cref{eq:Psiexp,eq:Upsexp} to calculate relations between the real and imaginary parts of the eigenvalues along the boundary, giving the expressions 
\begin{align}
 \text{Det}\left[\delta_{\alpha\mu} \delta_{\beta\nu}-\frac{1}{2}\sum_{\gamma\eta} q_{\alpha\gamma}\Psi^{\nu \eta}_{\mu\gamma}q^\dagger_{\eta\beta}\right]=0   \label{eq:Kdeteq}
\end{align}
and
\begin{align}
    \begin{aligned}
           \sum_\gamma(\omega\delta_{\alpha \gamma}-\mu_{\alpha\gamma} )q_{\beta\gamma}-\frac12 \sum_{\alpha'\gamma' \gamma}q_{\beta\gamma}q_{\alpha'\gamma'}\Upsilon^{\alpha'\gamma}_{\alpha\gamma'} &=\delta_{\alpha\beta}
    \end{aligned}\label{eq:allconds}
\end{align}
where Greek indices take values $1$ and $2$ corresponding to the price sector and the budget sector. One must solve for the auxiliary matrix $q$ in terms of the other terms to get a feasible expression in $\omega$ -- an easy analytical task for simple systems, but requiring numerical equation solving for complicated ones \cite{patil2023stability}. Comparing the exact eigenvalue spectra of the stability matrices given by \cref{eq:linneweqs} to the predictions for their boundaries coming from the correlations in \cref{eq:Psiexp,eq:Upsexp} in \cref{fig:spectravsmodel}, we see that we can get determine the stability of the system from the boundary of the spectrum in the complex plane. 

In the phase diagram in \cref{fig:spectravsmodel}, the dashed black line corresponds to this prediction from the boundaries. The red dotted line on the other hand corresponds to a prediction arising from scaling arguments -- we know that the bulk of the spectrum crossing into instability is centred around the value $\lambda_p=\kappa\log y$, and we wish to estimate the spread of this bulk. In the simplest case of a matrix with $N$ independent identically distributed variables with variance $\sigma^2/N$, it is known \cite{girko1985circular} that the radius of this bulk would be $\sigma$. For a matrix where values on opposite sides of the diagonal have correlation $\tau$, we also know \cite{sommers1988spectrum} that the resulting elliptical eigenvalue spectrum has its major and minor axes given by $\sigma(1\pm\tau)$.  As our system deals with block instead of individuals elements, we estimate the ``variance'' of each block as the variance of its modes. Each block has terms correlated in a way dictated by the matrix $\Psi$, and thus the eigenvalues of this matrix are the variances of the different independent modes of the block. The sum of these eigenvalues, or the trace of the covariance matrix, thus gives the overall variance of the block. This is simply the square root of the trace of the covariance matrix. We argue that we can predict the critical value of the production $y$ at which the transition must be occurring roughly by solving 
\begin{align}
     \log y_c  +\frac{\sqrt{{\rm Tr}[\Psi_0]}}{\sqrt{2N}}=0\label{eq:scalingargs}
\end{align} for $y_c$, where $\Psi_0$ is defined in Appendix \ref{sec:blockterms}. These predictions more accurate for systems with large $N$, as both the random matrix results and the assumption that the equilibrium state of is uniform rely on having sufficiently large systems. Note that for large $N$ we expect that $1 - y_c \propto N^{-1/2}$, independently of $\kappa$. Our economy is therefore not more stable when prices adapt faster to supply-demand imbalances, but benefit from having more agents, as fluctuations are reduced.  Note that  different choices for $\gamma$ and $\Gamma$ would not change the qualitative picture above, but only shift the transition to $\Gamma-y_c\propto \gamma N^{-1/2}$.

An important -- and somewhat unexpected -- conclusion emerging from the computation above is the following: the appearance of inequalities in our model is a collective, dynamical phenomenon. The structure of the eigenvector corresponding to the dynamical instability results from a subtle interplay between all the coefficients $J_{ij}$ and does not trivially correlate with individual properties, such as the average desirability of products.

\section{Discussion}
\subsection{Persistence of inequality and its solutions\label{sec:taxes}}

In an effort to understand the inequalities and what maintains them, we try intervening into this closed system with external injections of wealth/helicopter money \cite{friedman1969optimum}, redistribution of wealth (taxation), and altering preferences of different goods. It has been shown previously that providing money directly to the parts of society most stricken by poverty is a highly efficient means of combating inequality \cite{buiter2014simple}. We wish to see how this translates into our model. 

\begin{figure}
    \centering
    \hspace{0cm}\includegraphics[width=1\linewidth]{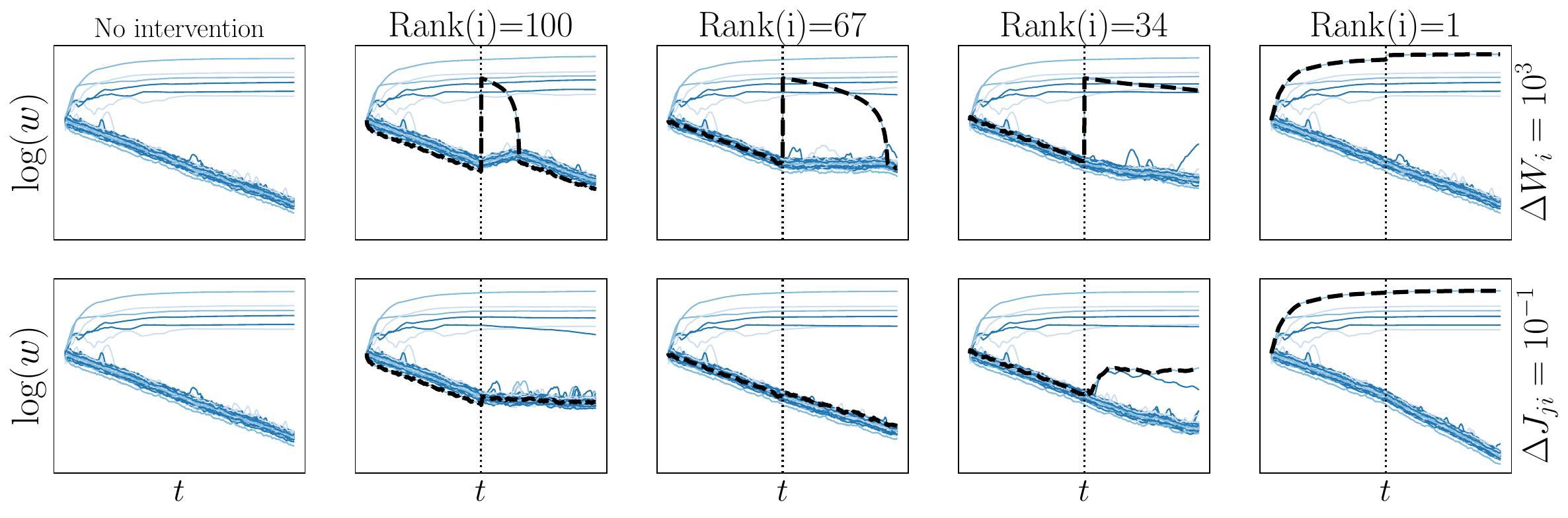}
    \caption{Increasing either the wealth of an agent or the desirability of their good produces a ripple effect through the system of $N=100$ people running for $T=1000$ time steps, with $y=0.75$. \textbf{Top}: `Helicopter Money'. We provide a one time lump-sum $\Delta w=1000$ units to rank $i$ agent at $t=500$. \textbf{Bottom}: Good Desirability. The various utilities of all agents $j \neq i$ for a particular person's good $i$ are increased uniformly by $\Delta J_{ji}=0.1$, also at $t=500$. For both types of interventions, we compare the wealth time series of a control system without intervention to 4 different systems in which intervention measures are made to the individuals with wealth ranks $i$ equal to $100$, $67$, $34$, and $1$ respectively. }
    \label{fig:helicopter}
\end{figure}
\begin{figure}
    \centering
    \includegraphics[width=\linewidth]{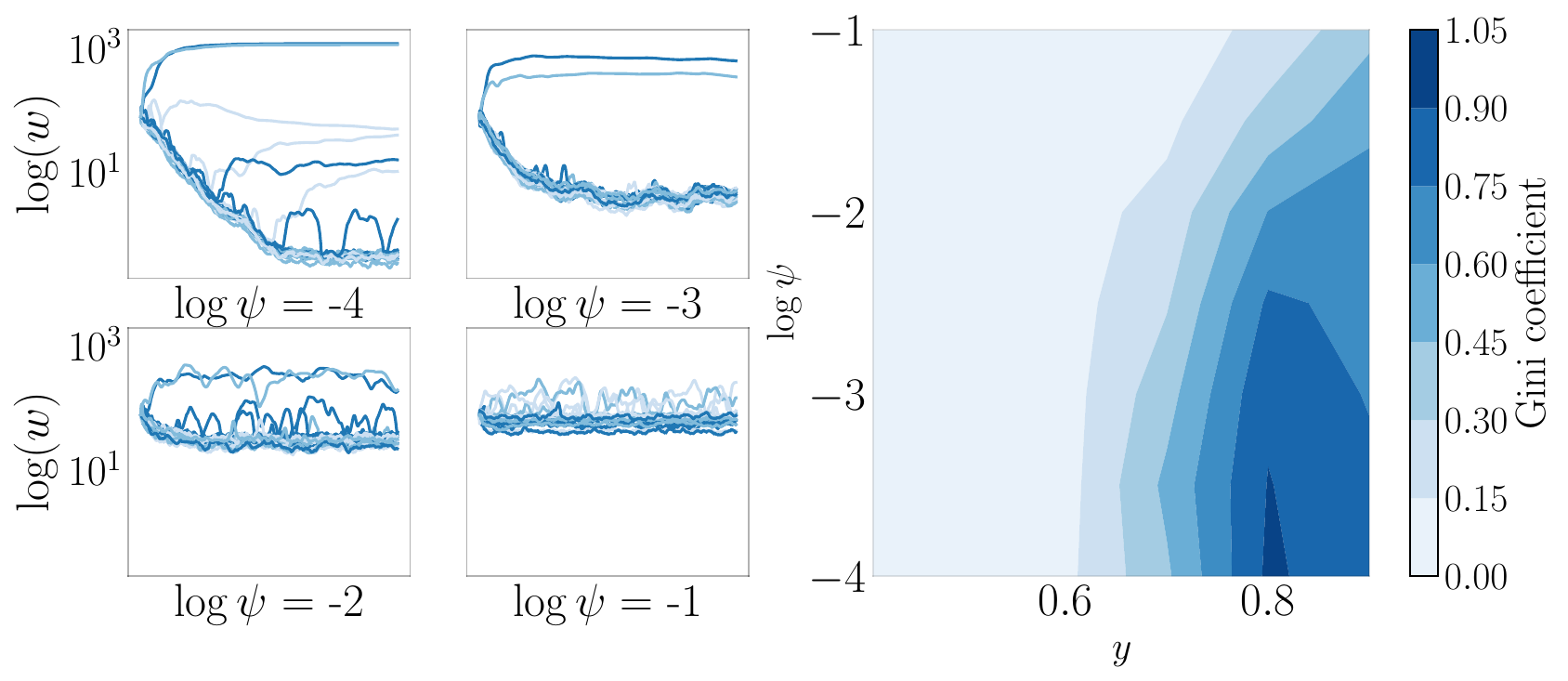}
    \caption{\textbf{Wealth Tax}: We redistribute a fraction $\rho$ of individual wealth every time step for a system with $N=100$ and $T=1000$. \textbf{Left}: For $y=0.8$, we see the wealth dynamics of some agents of this system for different values of the tax rate $\varphi$. For higher taxation, we find much lower inequalities in the model. \textbf{Right}: Phase diagram of the Gini coefficient as a function of quantity produced $y$ and the tax rate per unit time step $\varphi$ (in log base 10).}
    \label{fig:taxes}
\end{figure}
From the first row of graphs in \cref{fig:helicopter} we infer that providing a flat sum of money to a poor agent alleviates poverty across the board for a limited amount of time as it creates extra demand for everyone's goods, and spreads the wealth around. However, the system has a strong tendency to funnel wealth to the best ranking individuals, and making this resolution only short-lived. On the other hand, the same money given to wealthier individuals has no effect whatsoever, as expected since they do not use their wealth in any way at this point. 

In the second row, we show what happens if we take an agent $i$ and permanently increase how everyone else perceives their product, due to for example improved quality, external circumstances making their products necessary, or state subsidies. In the context of our model, this is implemented as $J_{ji} \to J_{ji} + \Delta J_{ji}$, $\forall j \neq i$. Given our analysis in \cref{fig:centcorrs} showing that wealth depends heavily on people's preferences, we expect some long term effect. If agent $i$ belong to the poor class it causes an actual difference in the final wealth ranking of said agent. Interestingly, it also leads to higher demand for all other products, causing price deflation to almost stop. However, if the same is done to an already rich agent, it forces others to spend more money on their product, further increasing inequalities. Intermediate cases lead to either no change at all (e.g. $i=67$), or to the agent getting out of the poverty trap without changing the fate of the rest of the population (see $i=34$).  

In \cref{fig:taxes}, we investigate the role of a wealth tax applied at each time step, the proceeds of which being redistributed equally between all agents. The first interesting effect is that price deflation stops after a while and the system becomes stationary. The gap between the rich class and the poor class is, as expected, a decreasing function of the tax rate $\varphi$. For $y=0.8$, the ratio of wealth between the richest and the poorest is $\sim 10^6$ for $\varphi = 0.01\%$ per time step and $10$ for $\varphi = 1 \%$ per time step. As can be seen in \cref{fig:taxes} (right), the inequality transition is smoothed out for large $\varphi$, with a Gini coefficient that does not exceed $0.65$ when $\varphi = 1 \%$. 

\subsection{Alternative dynamics\label{sec:altdyn}}

In our model, the inability of the system to coordinate around a set of prices that allow markets to clear so that everyone finish their goods causes instabilities and deflation. In a system normalised by average price, this would look like a world where a few individuals are able to make immense amounts of wealth initially, which then continues to grow multiplicatively. The rest of the population of such a society struggles to keep any amount of wealth, living paycheck to paycheck. However, as far as our model is concerned, these individuals are unaffected by the richer class, which continuous to accumulate wealth without being able to use it to consolidate power.

One of the factors that helps stabilise the system and bring about equality in the low production phase is the attempt at maximising the value for money $J/p$. If the variance of the utilities $J$ were high enough that it did not permit an equilibrium where prices were narrowly distributed around $p^\star$, the system would get more unstable, which intuitively fits into what we understand from May's limit \cite{may1972will}. A system where people do not consider the price at all and just buy objects with higher utilities would be a more extreme example of a similarly unstable system, see \cref{tab:table_compar}.
\begin{table}[t]
    \centering
    \begin{tabular}{|c|c|c|c|}
\hline
    Model & Stability  & Gini & $f$ rich \\ [1ex] 
    \hline 
    Normal dynamics & Unstable & 0.96 & 0.11\\
    Wealth in price update \cref{eq:wealth_in_priceupd} & Stable & 0.27 & 0.01\\
    Production update \cref{eq:Yadjusteqs} & Stable & 0.64 & 0.01\\
    Price independent purchases & Unstable & 0.97 & 0.06\\ [1ex] 
    \hline
\end{tabular}
    \caption{A comparison between different choices for the dynamics of the agents for a system with $N=100$ people, producing initially $y_0=1$ i.e. exactly enough goods for everyone, running for $T=10000$ time steps, and price update rate $r_p=0.01$. For the model with production updates, $r_y=r_p/2$. For this relatively high value of production, each system displays at least one ``elite'' agent who a few orders of magnitude more wealth than the common populace, though the Gini coefficient of the systems differ, showing different fates of the poorer class.}
    \label{tab:table_compar}
\end{table}
On the other hand, consider shifting the equilibrium condition such that instead of balancing sales and production, we weigh sales by the wealth of the client, which is to say
\begin{align}
    \frac{d p_i}{dt} &= \frac{\kappa p_i}{Ny}  (\sum_jT_{ji}\frac{b_j}{yp_i} - Ny). \label{eq:wealth_in_priceupd}
\end{align}
This means that sellers will increase price when their buyers can afford higher prices. This is not such an unnatural deviation from economic principles as the value of money reduces with income \cite{schultz1935interrelations}, and luxury goods are often more expensive not because of a high demand or low supply, but purely because their clientele can afford them. 

This modification can come with another variation, only allowing adding a weight to the sale for richer customers, and not removing any weight for poorer ones, or where one compares a client's wealth to that of the seller's, instead of to the price of their good. Similarly to the case where we allow the production to be adjusted, these modifications to the equations limits the depths of poverty in the system but still maintains strong inequalities for high values of production. Due to the balance between excess production and excess wealth, very high values of production result only in the gradual increase of the wealth or number of super rich agents as production is increased. 



\section{Conclusion}

We built an agent based good-exchange model with deceivingly simple rules, but with two important specific features: (a) preferences are heterogeneous, (b) goods are not divisible and utility saturates beyond some level of consumption. There is no money creation and no investments in our primitive economy. Using a simple price update rule that counteracts any supply-demand imbalances, we have found that a market clearing equilibrium can only be reached when production is low enough. Beyond a critical value of production, market clearing is still in principle achievable but become dynamically unstable, leading to unsold goods, deflation and the appearance of strong wealth inequalities. Situations where economic equilibrium exists but is dynamically unreachable were also discussed in \cite{bonart2014instabilities, dessertaine2022out}. Such a situation appears to be generic, as illustrated by the famous ``bullwhip effect'' (or beer game) \cite{Sterman1989}, see also \cite{sharma2021good} for a model of coordination breakdown.

We find this overall scenario rather interesting, as human societies have featured various transitions in production, whether it be the neolithic transition to agricultural societies, or the industrial revolutions \cite{allen2009engels,greenwood1997third,ryckbosch2016economic}. There is evidence to suggest these transitions also led to more pronounced hierarchies of power and socioeconomic class \cite{bender2005roots,bogaard2019farming,powers2014evolutionary,price1995social}. While there are numerous social factors behind these involving power, capital, and vested interests of individuals who were already at unequal footings, we would like to put forward our model as \textbf{(a)} an interesting mathematical problem in itself, and \textbf{(b)} a toy framework that provides several insights into emergent inequalities and how to mitigate them, e.g. through a wealth tax. It also illustrates how markets can be in unstable or critical equilibrium, resulting in major instabilities from small changes. We have indeed found that when production can adapt as well (and not only prices), the system settles around the critical point, providing an interesting realisation of self-organised criticality \cite{bak2013nature}. Our model adds to a long list of examples where strongly interacting natural or socio-technical systems spontaneously settle close to a state of marginal stability, see e.g. \cite{may1972will, bak1993aggregate, biroli2018marginally, moran2019may, beggs2012being, moran2023temporal} and refs. therein.

Finally, note that in our simplified model, it is impossible to spend more than the cost of the $N$ different goods. The accumulation of wealth is thus pointless and does not lead to a huge utility gap between the rich class and the poor class. But clearly, the appearance of unspent wealth opens the door to a capitalistic society. In that respect, the possibility of investing unspent wealth to improve the desirability of one's good would lead to further inequalities -- although such investments could also have off-shoot effects on the overall quality of products in the society. This observation is obviously at the heart of the capitalistic tension: wealth can be argued to be vital for innovation, and for the well-being of the society as a whole. The burning unanswered question, however, is: how much inequality is too much? \cite{rawls1971atheory, venkatasubramanian2017much, bouchaud2020much}

\section*{Acknowledgements}

We would like to thank Fabi\'an-Aguirre L\'opez, Jean-Pierre Nadal, Jerôme Garnier-Brun, and Salma Elomari for their constructive inputs on the model, our analytical methods, and the draft.

\begin{widetext}
    \appendix

    \section{Dynamical Stability with Classical Utilities}
\label{sec:classical}
Consider the alternate definition of the matrix $T$ via the classical utility route \cref{eq:classical_eq} such that
\begin{align}
    \begin{aligned}
        \frac{dp_i}{dt}&=\frac{\kappa p_i}{Ny}(\sum_jJ_{ji}\frac{b_j}{p_i\overline{J}}-Ny)\\
        \frac{db_i}{dt}&=\frac{\kappa}{N}\sum_j\left(\frac{J_{ji}b_j}{\overline{J}}-\frac{J_{ij}b_i}{\overline{J}}\right),
    \end{aligned}
\end{align}
where we have systematically replaced $\sum_kJ_{jk}$ by $N \overline{J}$ for large $N$, where  $\overline{J} := \int {\rm d}J J \rho(J)$.

This would give a block structured stability matrix whose blocks are given by the expressions
\begin{align}
    \begin{aligned}
        \mathbb{A}_{ij}&=-\delta_{ij}\kappa\\
        \mathbb{B}_{ij}&=\frac{\kappa}{N} \frac{J_{ji}}{y\overline{J}}\\
        \mathbb{C}_{ij}&=0\\
        \mathbb{D}_{ij}&=- \frac{\kappa}{N} \delta_{ij}\sum_k\frac{J_{ik}}{\overline{J}}+\frac{\kappa}{N} \frac{J_{ji}}{\overline{J}}
    \end{aligned}
\end{align}
which depend solely on the distribution of $J$'s. Neglecting fluctuations, i.e. for $J_{ij}= \overline{J}$, the eigenvalues of this matrix are easy to compute, because $ \mathbb{C}=0$. The $\mathbb{A}$ sector yields $N$ eigenvalues equal $\lambda_b=-\kappa$, whereas the $\mathbb{D}$ sector gives $N-1$ additional eigenvalues equal $\lambda_b=-\kappa$ and one associated with inflation invariance, $\lambda_0=0$. 
This means that there is no instability in this model, for any value of $y$. The fact that good quantities are discrete in our model does create instabilities. 

    \section{Stability matrix terms\label{sec:blockterms}}
In order to obtain the terms of the stability matrix in the linearised equations
\begin{align}
\begin{aligned}
        \begin{pmatrix}    \Dot{\Vec{\delta p}}\\\Dot{\Vec{\delta b}}
    \end{pmatrix}&=\begin{pmatrix}
    \mathbb{A} &\mathbb{B} \\\mathbb{C} & \mathbb{D}
    \end{pmatrix}\begin{pmatrix}
    \Vec{\delta p}\\\Vec{\delta b}
    \end{pmatrix},
\end{aligned}\label{eq:vecsupp}
\end{align}
consider the perturbations of the dynamical equations
\begin{align}
    \begin{aligned}
        \frac{dp_i^\star}{dt}+\frac{d\delta p_i}{dt}&= \frac{\kappa p_i^\star}{Ny}  (\sum_jT_{ji} - Ny)+\frac{\kappa \delta p_i}{Ny}  (\sum_jT_{ji} - Ny)+\frac{\kappa p^\star_i}{Ny}  (\sum_{jk}\frac{\partial T_{ji}}{\partial p_k}\delta p_k +\sum_{jk}\frac{\partial T_{ji}}{\partial b_k}\delta b_k)\\
         \implies\frac{d\delta p_i}{dt}&= \frac{\kappa \delta p_i}{Ny}  (\sum_jT_{ji} - Ny)+\frac{\kappa p^\star_i}{Ny}  (\sum_{jk}\frac{\partial T_{ji}}{\partial p_k}\delta p_k +\sum_{jk}\frac{\partial T_{ji}}{\partial b_k}\delta b_k)\label{eq:lin_newpeq}
    \end{aligned}
\end{align}
and for the budgets
\begin{align}
    \begin{aligned}
         \frac{d \delta b_i}{dt} &= \frac{1}{N} \sum_j\left(\delta p_iT_{ji}-T_{ij}\delta p_j\right)+\frac{1}{N} \sum_{jk}\left(p_i^\star\frac{\partial T_{ji}}{\partial b_k}\delta b_k+p_i^\star\frac{\partial T_{ji}}{\partial p_k}\delta p_k-p_j^\star\frac{\partial T_{ij}}{\partial b_k}\delta b_k-p_j^\star\frac{\partial T_{ij}}{\partial p_k}\delta p_k\right)
    \end{aligned}\label{eq:lin_newweq}
\end{align}
Taking the Fermi-Dirac Heaviside as before, we would have
\begin{align}
    \begin{aligned}
        T_{ij}&=\left({1+e^{-\beta(J_{ij}-\frac{p_j^\star}{\overline{p}^\star}P_>^{(-1)}(\frac{b_i^\star}{\overline{p}^\star}))}}\right)^{-1}\\
        \frac{\partial T_{ij}}{\partial p_k}&=\frac{\partial T_{ij}}{\partial J_{ij}}\left(-\frac{\delta_{jk}}{\overline{p}^\star}P_>^{(-1)}(\frac{b_i^\star}{\overline{p}^\star})+\frac{p_j^\star}{N\overline{p}^{*2}}P_>^{(-1)}(\frac{b_i^\star}{\overline{p}^\star})-\frac{p_j^\star}{\overline{p}^\star\rho(P_>^{(-1)}(\frac{b_i^\star}{\overline{p}^\star}))}\frac{b_i^\star}{N\overline{p}^\star} \right)\\
        \frac{\partial T_{ij}}{\partial b_k}&=\delta_{ik}\frac{\partial T_{ij}}{\partial J_{ij}}\frac{p_j^\star}{\overline{p}^{*2}}\frac{1}{\rho(P_>^{(-1)}(\frac{b_i^\star}{\overline{p}^\star}))}
    \end{aligned}
\end{align} using the definition given in \cref{eq:newdefTijgen}. Now assuming that the variations in the equilibrium variables themselves isn't that important, i.e.  $b_i^\star=yp_j^\star=y\overline{p}^\star$ for all $i$ and $j$, we get, with $Q(y):=\rho(P_>^{(-1)}(y))$,
\begin{align}
    \begin{aligned}
        \dot{\delta p_i}&= \frac{\kappa \delta p_i}{Ny}\sum_j\left[T_{ji}-T_{ji}'P_>^{(-1)}(y) - y\right]+\frac{\kappa }{NyQ(y)}  \sum_jT_{ji}'\delta b_j+\frac{\kappa}{N^2y}\sum_{jk}T_{ji}'\left[P_>^{(-1)}(y)-\frac{y}{Q(y)}\right]\delta p_k\\
         \dot{\delta b_i} &=\sum_k\frac{\kappa\delta p_k}{N}\sum_j\Bigg[ T_{ij}'\left(\delta_{jk}P_>^{(-1)}(y)-\frac{P_>^{(-1)}(y)}{N}+\frac{y}{NQ(y)}\right)-T_{ji}'\left(\delta_{ik}P_>^{(-1)}(y)-\frac{P_>^{(-1)}(y)}{N}+\frac{y}{NQ(y)}\right)\\&\qquad\qquad\qquad  +\delta_{jk}T_{ij}\Bigg]-\frac{\kappa}{NQ(y)}\sum_j\left(\delta b_iT_{ij}'-\delta b_jT_{ji}'\right)&\qquad 
    \end{aligned}\label{eq:linneweqs}
\end{align}
which means
\begin{align}
    \begin{aligned}
        \mathbb{A}_{ij}&=\frac{\kappa}{Ny}\sum_k\left[\delta_{ij}(T_{ki}-T_{ki}'P_>^{(-1)}(y)-y)-\frac{1}{N}T'_{ki}\left(\frac{y}{Q(y)}-P_>^{(-1)}(y)\right)\right]\\
        \mathbb{B}_{ij}&=\frac{\kappa}{NyQ(y)}T_{ji}'\\
        \mathbb{C}_{ij}&=\frac{\kappa}{N}\sum_k\left[\frac{1}{N}\left(\frac{y}{Q(y)}-P_>^{(-1)}(y)\right)(T_{ik}'-T_{ki}')-(\delta_{ij}T_{ki}'-\delta_{jk}T_{ik}')P_>^{(-1)}(y)+\delta_{ij}T_{ki}-\delta_{jk}T_{ik}\right]\\
        \mathbb{D}_{ij}&=-\delta_{ij}\frac{\kappa}{NQ(y)}\sum_kT_{ik}'+\frac{\kappa}{NQ(y)}T_{ji}'
    \end{aligned}\label{eq:blockelements}
\end{align}
Now the averages of the matrices $T$ and $T'$ themselves over the choices for utilities $J$ are given by
\begin{align}
    \begin{aligned}
        \overline{T}=y,\qquad \overline{T}'=Q(y)
    \end{aligned}
\end{align}
which means that the stability matrix would have the following values in the absence of disorder
\begin{align}
    \begin{aligned}
        \overline{\mathbb{A}}&=-\delta_{ij}\frac{\kappa}{y}Q(y)P_>^{(-1)}(y)-\frac{\kappa}{Ny}(y-P_>^{(-1)}(y)Q(y))=(Na-\kappa)\mathbb{I}-a\mathbb{E}\\
        \overline{\mathbb{B}}&=\frac{\kappa}{NP_>^{(-1)}(y)Q(y)}=g\mathbb{E}\\
        \overline{\mathbb{C}}&=\delta_{ij}\kappa (y-P_>^{(-1)}(y)Q(y))-\frac{\kappa }{N}(y-P_>^{(-1)}(y)Q(y))=Nc\mathbb{I}-c\mathbb{E}\\
        \overline{\mathbb{D}}&=-\delta_{ij}\kappa+\frac{\kappa}{N}=-Nd\mathbb{I}+d\mathbb{E}
    \end{aligned}\label{eq:constmatgen}
\end{align}
where $\mathbb{I}$ is the identity matrix and $\mathbb{E}$ is a matrix of all $1$'s. The characteristic equation of this matrix is easy to write, as
\begin{align}
    \begin{aligned}
        \det[\mathbb{M}-\lambda\mathbb{I}]&=\det[(Na\mathbb{I}-\kappa\mathbb{I}-\lambda\mathbb{I}-a\mathbb{E})(-Nd\mathbb{I}-\lambda\mathbb{I}+d\mathbb{E})+g\mathbb{E}(\mathbb{c}E-Nc\mathbb{I})]\\
        &=\lambda(\lambda+\kappa)(\lambda+\kappa-Na)^{N-1}(\lambda+Nd)^{N-1}
    \end{aligned}
\end{align}
telling us that the eigenvalues of the stability matrix are centred around 
\begin{align}
    \lambda_p=-\frac{\kappa}{y}P_>^{(-1)}(y)  Q(y), \qquad {\rm and} \qquad \lambda_b=-\kappa
\end{align}
and two other eigenvalues $\lambda_0=0$ and $\lambda_1=-\kappa$. In the rest of the work, we take the distribution $\rho(J)=e^{-J}$, making the first bulk of eigenvalues to be at $\lambda_p=\kappa \ln y$, which behaves as $-\kappa (1- y)$ when $y\rightarrow 1$, resulting in the instability we find. 

More generally, if $\rho(J)$ behaves as $(J/J_0)^\mu$ when $J \to J_{\min} \geq 0$, one find that $\lambda_p$ behaves as $-\kappa (1+ \mu) (1-y)$  when $y\rightarrow 1$, independently of both $J_{\min}$, the minimum value of $J$ and $J_0$. This suggests that as small $J$ values become rarer, the economy is more stable, as intuitively expected. On the contrary, when there is an accumulation of unwanted goods (i.e. $-1 < \mu < 0$) the system is more prone to instabilities. Note that the case $\rho(J)=e^{-J}$ corresponds to $P(J \to 0)=1$, i.e., $\mu=0$.

\section{Matrix manipulations and correlations}

We rotate the $2N\times2N$ stability matrix to limit the correlations within $2\times2$ blocks, creating 
\begin{align}
\mathbb{R}_{ij}^{\alpha\beta}&=\sum_{\gamma\delta kl}V^{(i,\gamma)}_{k\alpha }\mathbb{L}_{kl}^{\gamma\delta}V^{(j,\delta)}_{l\beta}
\end{align}
where $V^{(i,\alpha)}_{j\beta}=\delta^{(2)}_{\alpha\beta}W^{(N)}_{ij}$, with the vector basis $W^{(n)}$ being an orthogonal basis of vectors of size $n$ that includes the vector $w_1=\frac{1}{\sqrt{n}}\begin{pmatrix}
    1&1&1&\dots&1
\end{pmatrix}^T$. The matrix $\mathbb{L}$ is the original stability matrix from \cref{eq:veceqdef}. The resulting matrix has correlations 
\begin{align}
    2N\overline{\delta \mathbb{R}_{ij}^{\alpha\beta} \delta\mathbb{R}_{kl}^{\alpha'\beta'}}=\Psi_{\alpha\beta}^{\alpha'\beta'}\delta_{ik}\delta_{jl}+\Upsilon_{\alpha\beta}^{\alpha'\beta'}\delta_{il}\delta_{jk}
    \label{eq:covariances}
\end{align}
The average values of the rotated matrix are located around the diagonal in the form
\begin{align}
\overline{\mathbb{R}}=\delta_{ij}\begin{pmatrix}
        \kappa \ln y&0\\\kappa y(1+\ln y)&-\kappa
    \end{pmatrix}
\end{align}
and the correlation matrices are $\Psi=2\kappa^2 \Psi_0/N$ and $\Upsilon= 2\kappa^2 \Upsilon_0/N$, with 
\begin{align}
    \begin{aligned}
        \Psi_0=&\begin{pmatrix}
            \begin{pmatrix}
                \frac{(\ln y)^2}{y^2} & 0\\\frac{(\ln y)^2}{y}&0
            \end{pmatrix}&\begin{pmatrix}
                0 & \frac{1}{y^4}\\0 & \frac{1}{y^3}
            \end{pmatrix}\\
             \begin{pmatrix}
                \frac{(\ln y)^2}{y} & 0\\2(\ln y)^2 &0
            \end{pmatrix}&\begin{pmatrix}
                0 & \frac{1}{y^3}\\0 & \frac{2}{y^2}
            \end{pmatrix}
        \end{pmatrix}y(\frac{\beta}{6}-y)+\begin{pmatrix}
            \begin{pmatrix}
                \frac{1}{y^2} & 0\\\frac{1}{y}&0
            \end{pmatrix}&\begin{pmatrix}
                0&0\\0&0
            \end{pmatrix}\\
            \begin{pmatrix}
                \frac{1}{y}&0\\2&0
            \end{pmatrix}&\begin{pmatrix}
                0&0\\0&0
            \end{pmatrix}
        \end{pmatrix}y(1-y)\\
        &+\begin{pmatrix}
            \begin{pmatrix}
                2\frac{1}{y^2} & 0\\2\frac{1}{y}&0
            \end{pmatrix}&\begin{pmatrix}
                0&0\\0&0
            \end{pmatrix}\\
            \begin{pmatrix}
                2\frac{1}{y}&0\\4&0
            \end{pmatrix}&\begin{pmatrix}
                0&0\\0&0
            \end{pmatrix}
        \end{pmatrix}y(\frac{1}{2}-y)\ln y
    \end{aligned}\label{eq:Psiexp}
\end{align}
and
\begin{align}
    \begin{aligned}
        \Upsilon_0=&\begin{pmatrix}
            \begin{pmatrix}
                \frac{(\ln y)^2}{y^2} & 0\\\frac{(\ln y)^2}{y} &0
            \end{pmatrix}&\begin{pmatrix}
                0 & 0\\-\frac{\ln y}{y^2} & 0
            \end{pmatrix}\\
             \begin{pmatrix}
                \frac{(\ln y)^2}{y}& -\frac{\ln y}{y^2}\\(\ln y)^2 &-\frac{\ln y}{y}
            \end{pmatrix}&\begin{pmatrix}
                0 & 0\\-\frac{\ln y}{y} & \frac{1}{y^2}
            \end{pmatrix}
        \end{pmatrix}y(\frac{\beta}{6}-y)+\begin{pmatrix}
            \begin{pmatrix}
                \frac{1}{y^2} & 0\\\frac{1}{y}&0
            \end{pmatrix}&\begin{pmatrix}
                0&0\\0&0
            \end{pmatrix}\\
            \begin{pmatrix}
                \frac{1}{y}&0\\1&0
            \end{pmatrix}&\begin{pmatrix}
                0&0\\0&0
            \end{pmatrix}
        \end{pmatrix}y(1-y)\\
        &+\begin{pmatrix}
            \begin{pmatrix}
                2\frac{1}{y^2}\ln y & 0\\2\frac{1}{y}\ln y&0
            \end{pmatrix}&\begin{pmatrix}
                0&0\\-\frac{1}{y}&0
            \end{pmatrix}\\
            \begin{pmatrix}
                2\frac{1}{y}\ln y&-\frac{1}{y}\\2\ln y&-1
            \end{pmatrix}&\begin{pmatrix}
                0&0\\-1&0
            \end{pmatrix}
        \end{pmatrix}y(\frac{1}{2}-y)
    \end{aligned}\label{eq:Upsexp}
\end{align}
where one set of indices (upper or lower) on $\Psi^{\alpha'\beta'}_{\alpha\beta}$ or $\Upsilon^{\alpha'\beta'}_{\alpha\beta}$  indicate which $2\times2$ block we're in, and the other set indicates which element of the block we choose.

\end{widetext}
\bibliography{apssamp}

\end{document}